\documentclass[aps,prd,twocolumn,nofootinbib]{revtex4-1}
\usepackage{epsfig}
\usepackage[colorlinks,linkcolor=blue,anchorcolor=blue,citecolor=blue,urlcolor=blue,breaklinks=true]{hyperref}
\usepackage{graphics}
\usepackage{slashed}
\usepackage{color}
\usepackage{amsfonts}
\usepackage{amsmath}
\usepackage{extarrows}
\usepackage{amssymb}

\begin{document}
\author{Chao Shi$^{1,2,4}$}
\author{Yi-Lun Du$^{1}$}
\author{Shu-Sheng Xu$^{2,4}$}
\author{Xiao-Jun Liu$^{1}$}\email{Email:liuxiaojun@nju.edu.cn}
\author{Hong-Shi Zong$^{2,3,4}$}\email{Email:zonghs@nju.edu.cn}

\address{$^{1}$ Key Laboratory of Modern Acoustics, MOE, Institute of Acoustics, and Department of Physics, Collaborative Innovation Center of Advanced Microstructures, Nanjing University, Nanjing 210093, China}
\address{$^{2}$ Department of Physics, Nanjing University, Nanjing 210093, China}
\address{$^{3}$ Joint Center for Particle, Nuclear Physics and Cosmology, Nanjing 210093, China}
\address{$^{4}$ State Key Laboratory of Theoretical Physics, Institute of Theoretical Physics, CAS, Beijing 100190, China}

\title{Continuum study on QCD phase diagram through an OPE-modified gluon propagator}
\begin{abstract}
Within the Dyson-Schwinger equations (DSEs) framework, a  gluon propagator model incorporating quark's feedback through operator product expansion (OPE) is introduced  to investigate the QCD  phase diagram in the temperature--chemical-potential ($T-\mu$) plane. Partial restoration of chiral symmetry at zero temperature and finite temperature are both studied, suggesting a first order phase transition point on the $\mu$ axis and a critical end point at $(T_E,\mu_E)/T_c = (0.85,1.11)$, where $T_c$ is the pseudo-critical temperature. In addition, we find the pseudo-critical line can be well parameterized with the curvature parameter $\kappa$  and a consistent decrease in $\kappa$ with more of gluon propagator distributed to quark's feedback.

\bigskip


\bigskip

\noindent PACS Number(s): 25.75.Nq, 11.30.Rd, 11.10.Wx, 12.38.Lg

\end{abstract}
\maketitle

\vbox{}

\section{Introduction}\label{sec:intro}
The universe went through a quark epoch approximately $10^{-12}$ seconds after the Big Bang. Nowadays, the nucleus-nucleus collisions  at RHIC and LHC with high center of mass energy can reproduce such a state  known as the quark gluon plasma (QGP) \cite{Adams:2005dq,Shuryak:2008eq}. It consists of unbound quarks/gluons and behaves as nearly perfect fluid \cite{Shuryak:2003xe,Zajc:2007ey} with very small viscosity. While progresses are being made in studying the QGP concerning the high temperature ($T$) and low chemical potential ($\mu$) region in the QCD phase diagram, little is known about the territory with higher $\mu$. Hence, RHIC is planning a beam energy scan program phase II (BES II)  based on the  BES I completed in 2014 \cite{Odyniec:2015iaa,McDonald:2015tza}. With statistical errors largely reduced, strong conclusions on QCD phase transition boundary and the critical end point (CEP) are hopefully to be drawn.

On the theoretical side, with finer lattices and physical quark masses, lattice simulations observed the analytical crossover behavior at $T\ne 0, \mu = 0$ and investigated various thermodynamic quantities of the QGP \cite{Bazavov:2011nk,Borsanyi:2010bp,Bazavov:2014pvz}. However, its extrapolation to $\mu \ne 0$ is a yet unsolved problem due to the notorious sign problem \cite{Troyer:2004ge,deForcrand:2010he,Endrodi:2011gv}. Therefore, alternative approaches to the QCD phase diagram like (P)NJL models \cite{Costa:2008yh,Costa:2008gr}, quark-meson models \cite{Nickel:2009wj,Skokov:2010uh,Ayala:2014jla} and the Dyson-Schwinger equations (DSEs) method \cite{Qin:2010nq,Fischer:2012vc,Xin:2014ela,Shi:2014zpa}, could provide valuable insight at present \cite{Weise:2012yv}.

In this work, we will resort to DSEs, which is a continuous non-perturbative approach that describes QCD's several important features, e.g., dynamical chiral symmetry breaking (DCSB) and confinement \cite{Bashir:2012fs}. It has been employed in extensive study on the QCD phase diagram. For instance, chiral phase restoration was studied over the $T-\mu$ space and generally speaking, the existence of CEP is suggested, in consistent with most model predictions. It is further supplemented by the investigation of certain phases, e.g., sQGP at high temperature \cite{Gao:2014rqa,Gao:2015kea} and color superconductivity at  low temperature \cite{Muller:2013pya}. The effect of chiral imbalance on the QCD phase structure is also studied, extending the  phase diagram to $T-\mu-\mu_5$ space \cite{Xu:2015vna,Wang:2015tia}.

As an  infinite tower of equations, DSEs  always require truncation schemes in practice.  For example, the quark's DSE, namely the quark gap equation, has two unknown ingredients: quark-gluon vertex and gluon propagator. For the quark-gluon vertex, commonly used are: i) Rainbow truncation, namely the bare vertex which had been widely used in combination with Ladder truncation in bound state problems. ii) Ball-Chiu (BC) ansatz \cite{Ball:1980ay} and its modified versions that concern the Abelian and non-Abelian dressing effects \cite{Fischer:2009wc}. iii) Ball-Chiu ansatz plus a dressed-quark anomalous chromomagnetic moment term \cite{Gao:2014rqa}. In spite of that the latter two vertexes are more refined, Rainbow truncation suffices to give qualitative descriptions of the QCD  phase diagram in almost all aspects. Therefore  we will use it throughout this work for simplicity.

With the Rainbow truncation, our main focus in the work will be on the other ingredient of the quark self-energy, the gluon propagator. A popular choice is to directly generalize models determined in hadron physics, e.g., separable model \cite{Burden:1996nh,Blaschke:2000gd}, Maris-Tandy model \cite{Maris:1999nt}, Qin-Chang model \cite{Qin:2011dd} and etc \footnote{Our work will be based on these bottom-up-scheme models which are determined by fitting hadron properties \cite{Binosi:2014aea}. The other scheme, top-down scheme, which aims to perform an ab initio computation of the gauge-sector DSEs can be tracked from \cite{Fischer:2003rp, Aguilar:2008xm}.}, to the finite temperature case \cite{He:2009hv,Qin:2010nq}.  However, flaws in these generalized models are apparent. They receive no feedback from quarks and don't evolve with temperature or chemical potential, therefore don't meet the  requirements of QCD in essence. A specific example is the first order chiral phase transition at low temperature and high density.  There the gluon propagators in Nambu-Goldstone phase and in Wigner phase should be different and a discontinuous change is expected.  In face of this situation, authors of \cite{Fischer:2011mz,Fischer:2012vc} incorporate quark's feedback into the gluon propagator by considering contribution of quark loops  in gluon's DSE. Nevertheless, the quenched part of gluon propagator relies on analyzing and fitting the lattice data.   

In this paper, we will investigate an alternative treatment based on the operator product expansion (OPE), which provides an explicit form for quark's feedback on gluon self-energy in terms of local quark condensates \cite{Steele:1989ik,Jiang:2011up}. In this way, we derive a modified gluon propagator model and the consequent QCD  phase diagram is studied within DSEs framework. Since the extraction of quark's feedback on gluon remains an open question, our model study will hopefully help us gain useful insights. 
	
This paper is organized as follows. In Sec.~\ref{sec:dsesandgp} we introduce quark's gap equation and the truncation scheme. Then a gluon propagator model is  derived from gluon DSE with the help of OPE. With this model, we study the transition behavior of QCD on the $T-\mu$ plane in remaining sections, where the case of $T=0,\mu \ne 0$ is discussed in Sec.~\ref{sec:t0} and $T \ne 0,\mu \ne 0$ is studied in Sec.~\ref{sec:tu}. Finally we summarize our result and give the conclusions in Sec.~\ref{sec:summary}.

\section{Quark gap equation and gluon propagator model}
\label{sec:dsesandgp}
To study the QCD chiral phase transition, we employ the Dyson-Schwinger equations formalism, in which the quark gap equation at finite temperature and chemical potential can be written as

\begin{eqnarray}
\label{eq:gapeq}
\hspace{-5mm} [ G(\vec{p},\tilde{\omega}_n)]^{-1}&=&[G^0(\vec{p},\tilde{\omega}_n)]^{-1}+T\sum_{l=-\infty}^\infty\int\frac{d^3q}{(2\pi)^3}
\nonumber\\
&&\hspace*{-6mm}\times \, \left[g^2D_{\mu \nu}(\vec{p}\!-\!\vec{q},\tilde{\omega}_n\!\!-\!\tilde{\omega}_l)\frac{\lambda^a}{2}\gamma_{\mu}G(\vec{q},\tilde{\omega}_l)\Gamma_{\nu}^a \right],
\end{eqnarray}
where the superscript \emph{0} refers to free propagators. $\tilde{\omega}_n$=$(2n+1)\pi T+i\mu$ and the color index in gluon propagator $D_{\mu\nu}$ has been contracted. $\lambda^a$ are the Gell-Mann matrices and $\Gamma_\nu^a$ is the full quark gluon vertex. Here we have set all renormalization constants to one, since we will use gluon models that are heavily suppressed in ultraviolet region, rendering the integral in quark self-energy convergent. In this sense, the $g^2$ here is not a running coupling constant in the sense of the renormalization group but rather an effective coupling and therefore has no medium dependence. We use Landau gauge here, which is  a fixed point of the renormalization group and therefore widely used in DSEs studies \cite{Roberts:2000aa}. The quark propagator can further be decomposed as
\begin{eqnarray}
\label{eq:GABC}
G^{-1}(\vec{p},\tilde{\omega}_n;T,\mu)&=&i\vec{\gamma}\cdot\vec{p}A(\vec{p}^{~\!2},\tilde{\omega}_n^2;T,\mu) 
\nonumber\\
&&\hspace*{-15mm}+i\gamma_4\tilde{\omega}_nC(\vec{p}^{~\!2},\tilde{\omega}_n^2;T,\mu)+B(\vec{p}^{~\!2},\tilde{\omega}_n^2;T,\mu).
\end{eqnarray}
For the free quark propagator $G^0(\vec{p},\tilde{\omega}_n)$, scalar functions $A=1$, $B=m$ and $C=1$, where $m$ is the current quark mass. Rainbow truncation has been  popular in meson study because its combination with Ladder truncation preserves the axial-vector Ward-Takahashi identity \cite{Maris:1997tm}. And in our case, as far as we know, no existing complicated vertexes bring  qualitative changes to the  phase diagram. So for simplicity, we will employ the rainbow truncation through out this work, namely

\begin{equation}
\label{eq:rainbow}
\Gamma^a_\nu(p,q)=\frac{\lambda^a}{2}\gamma_\nu .
\end{equation}
 In this way, we are left with the gluon propagator which is undetermined. Generally, it can also be expressed through two scalar functions $D_T$ and $D_L$

\begin{align}
 D_{\mu \nu}(\vec{k},\Omega_l)=P_{\mu \nu}^T (\vec{k},\Omega_l) D_T(\vec{k}^2,\Omega_l^2)
\nonumber\\
 +P_{\mu \nu}^L(\vec{k},\Omega_l) D_L(\vec{k}^2,\Omega_l^2),
\end{align}
with $P_{\mu \nu}^L$ and $P_{\mu \nu}^T$ being longitudinal and transverse projection operators respectively
\begin{eqnarray}
P_{\mu \nu}^T (k)&=&(1-\delta_{\mu 4})(1-\delta_{\nu 4})(\delta_{\mu \nu}-\frac{k_\mu k_\nu}{\vec{k}^2}) \\
P_{\mu \nu}^L (k)&=&P_{\mu \nu}(k)-P_{\mu \nu}^T (k) .
\end{eqnarray}
$\Omega_l=2 l \pi T$ is the boson Matsubara frequency. Normally, one can now resort to the aforementioned models, e.g., Qin-Chang model for $D_T$ and $D_L$. However, since we are trying to incorporate quark's feedback, further consideration is needed. Let's start with the case at zero temperature and density.

\begin{figure}
\includegraphics[width=0.47\textwidth]{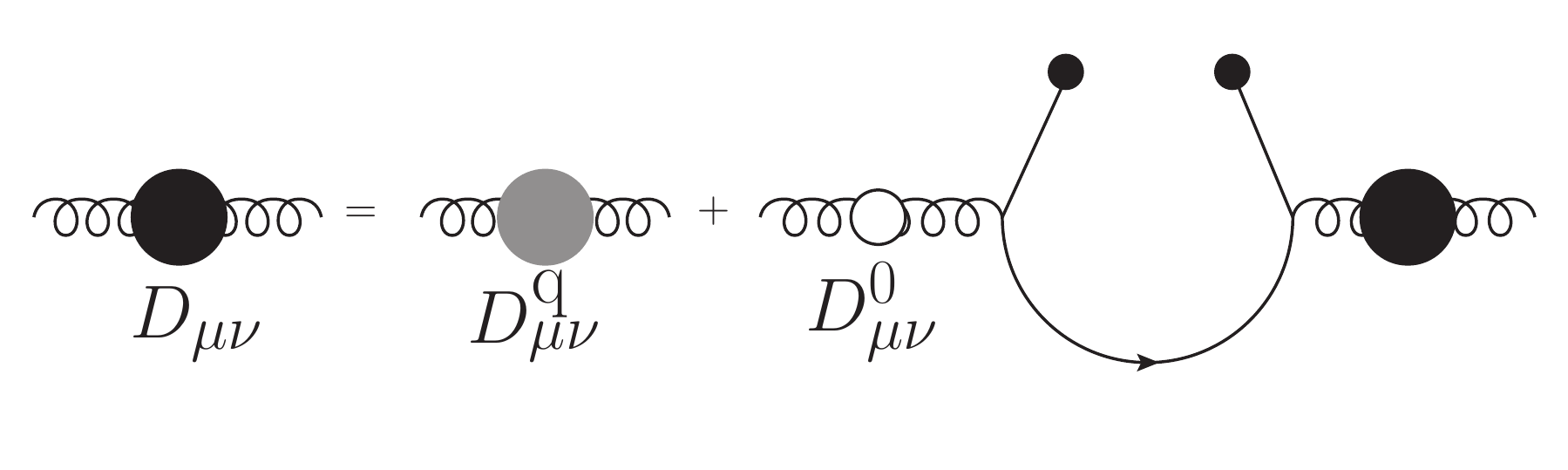}
\caption{Gluon DSE with a vacuum polarization term which contains local quark condensate, see Eq.~(\ref{eq:pimunu}) and Eq.~(\ref{eq:gluonDSE}).}
\label{fig:gluon_DSE}
\end{figure}

As we mentioned in Sec.~\ref{sec:intro}, extracting quark's feedback from gluon propagator is tricky. Inspired by QCD sum rule \cite{Shifman:1998rb}, authors of \cite{Jiang:2011up} suggested a relatively simple way as follows. As we know, in the OPE framework, the current-current correlation function can be expressed through various local scalar operators' vacuum expectation values, namely, vacuum condensates. These vacuum condensates characterizing the non-perturbative feature of QCD  are treated as parameters in the QCD sum rules, while they can be calculated elsewhere \cite{McNeile:2012xh}, including DSEs \cite{Maris:1997tm,Zong:2002ve}. For the gluon propagator, the gluon self-energy contains quark condensate, which is the lowest dimension vacuum condensate generated by quarks.  The gluon vacuum polarization tensor involves the term \cite{Steele:1989ik,Jiang:2011up}

\begin{eqnarray}
\label{eq:pimunu}
\Pi^{\textrm{Q}}_{\mu \nu}(k)&=&-g^2\int d^4(y-z)\int \frac{d^4q}{(2 \pi)^4}e^{i(p-q)\cdot (y-z)}\nonumber \\
\  \ & & \times \textrm{tr}\biggl{[}\gamma_\mu \frac{1}{i \slashed{q}+m}\gamma_\nu \langle \bar{\psi}(y)\psi(z) \rangle \biggl{]} \nonumber \\
&=&P_{\mu \nu}(k)k^2 \Pi^{\textrm{Q}}(k^2)\nonumber \\ 
&=&-P_{\mu \nu}(k) \frac{g^2 m \left \langle  \bar{\psi} \psi \right \rangle }{3k^2}+...
\end{eqnarray}
where $m \left \langle  \bar{\psi} \psi \right \rangle=m_u\langle \bar{\psi}\psi \rangle_u+m_d\langle \bar{\psi}\psi \rangle_d$ and the ellipsis represents terms of higher orders in $m^2/k^2$ which are neglected. The superscript \emph{Q} stands for \emph{Quark}. Now, we can extract from the full gluon propagator a quark-unaffected part $D^{\textrm{q}}$, where \emph{q} stands for \emph{quenched}. Then the full gluon propagator is divided into two parts,
\begin{eqnarray}
\label{eq:D1}
D_{\mu \nu}(k)&=&P_{\mu \nu} D(k^2)\\ 
\label{eq:D2}
&=&P_{\mu \nu} (D^{\textrm{q}}(k^2)+D^{\textrm{Q}}(k^2)) .
\end{eqnarray} 
Accordingly, with the DSE for gluon propagator we have
\begin{eqnarray}
\label{eq:gluonDSE}
D_{\mu \nu}(k)=D^{\textrm{q}}_{\mu \nu}(k)+D^0_{\mu \rho}(k) \Pi_{\rho \sigma}^{\textrm{Q}}(k) D_{\sigma \nu}(k) ,
\end{eqnarray}
which is diagrammatically shown in Fig.~\ref{fig:gluon_DSE}. With Eqs.~(\ref{eq:pimunu},\ref{eq:D2},\ref{eq:gluonDSE}), we have:
\begin{eqnarray}
\label{eq:Dk2}
D(k^2)&=&\frac{D^{\textrm{q}}(k^2)}{1+\frac{\displaystyle g^2 m\langle \bar{\psi}\psi \rangle_0}{3k^4}} \nonumber \\
&\approx &\frac{D^{\textrm{q}}(k^2)}{\displaystyle 1+\frac{\langle \bar{\psi}\psi \rangle_0}{ \Lambda^3}} ,
\end{eqnarray}
where the subscript \emph{0} refers to $T=0$ and $\mu=0$. Here we introduce the momentum scale $\Lambda$ as in \cite{Jiang:2011up}, which  absorbs constants $m$, $g$ and the momentum $k$ and serves as a parameter in our model. With such simplification, gluon propagator remains finite in the infrared region and the ultraviolet region won't be affected since $D^{\textrm{q}}(k^2)$ will be heavily ultraviolet-suppressed.

Then we extend Eq.~(\ref{eq:Dk2}) to finite temperature and chemical potential via $k \to k_l=(\vec{k},\Omega_l)$ and $\langle \bar{\psi} \psi \rangle_0 \to \langle \bar{\psi} \psi \rangle_{T,\mu}$, so

\begin{eqnarray}
\label{eq:Dk2O21}
D(\vec{k}^2+\Omega_l^2)= \frac{D^{\textrm{q}}(\vec{k}^2+\Omega_l^2)}{\displaystyle 1+\frac{\langle \bar{\psi}\psi \rangle_{T, \mu}}{ \Lambda^3}}.
\end{eqnarray}
Note that here is an implicit approximation $D_L=D_T$, which actually doesn't hold at finite temperature as shown by lattice simulation \cite{Cucchieri:2007ta,Silva:2013maa}. However, for a sketchy study of quark's feedback on chiral phase diagram, we will continue to use this approximation following earlier studies \cite{He:2009hv,Qin:2010nq}.

To specify the function $D^{\textrm{q}}(k^2)$, we will employ the Qin-Chang model as the full gluon propagator at zero temperature and density, 

\begin{align}
\label{eq:QCmodel1}
 g^2 D_{\mu \nu}(k^2)&= {\cal G} (k^2) P_{\mu\nu} \\
\label{eq:QCmodel2} 
\mathcal{G}(k^2)&=\frac{8\pi^2}{\omega^4}D e^{-\frac{k^2} {\omega^2}} ,
\end{align}
where the parameters $D$ and $\omega$ are determined in hadron physics. $D$ characterizes the interaction strength and $\omega$ controls the confinement length. In Rainbow-Ladder truncation, the ground state pseudoscalar and vector-meson observables, like mass and electro-weak decay constant, are roughly constant while $Dw=(0.8\textrm{GeV})^3$ with $\omega \in [0.4,0.6] \textrm{GeV}$. Therefore  the parameters are not completely constrained by hadron physics: a change in D can be compensated by an alteration of $\omega$. Qin-Chang model qualitative agrees with modern DSEs and lattice studies in gluon propagator's infrared region, e.g., it gives typical value for the gluon screening mass \cite{Qin:2011dd}.\footnote{
Qin-Chang model improves upon an alike model: Maris-Tandy model 
\begin{align}
\label{eq:MTmodel}
\mathcal{G}(k^2)=\frac{4\pi^2}{\omega^6}D k^2 e^{-\frac{k^2}{\omega^2}} 
\end{align}
in the deep-infrared region of gluon propagator.}
So with Eqs.~(\ref{eq:D1},\ref{eq:Dk2},\ref{eq:QCmodel1}), we have
\begin{align}
\label{eq:Dq2}
D^q(k^2)=\frac{ {\cal G} (k^2)}{g^2} \left ( \displaystyle 1+\frac{\langle \bar{\psi}\psi \rangle_{0}}{ \Lambda^3} \right ).
\end{align}
Substituting it into Eq.~(\ref{eq:Dk2O21}), we finally arrive at the OPE-modified model

\begin{eqnarray}
\label{eq:Dk2O2}
g^2 D(\vec{k}^2+\Omega_l^2)=\mathcal{G}(\vec{k}^2+\Omega_l^2)\frac{\displaystyle 1+\frac{\langle \bar{\psi}\psi \rangle_0}{\Lambda^3} }{\displaystyle 1+\frac{\langle \bar{\psi}\psi \rangle_{T,\mu}}{\Lambda^3} }. 
\end{eqnarray}
Apparently, the form of $g^2 D(\vec{k}^2+\Omega_l^2)$ changes as $\langle \bar{\psi}\psi \rangle$ evolves through the $T-\mu$  plane. At $T=0$ and $\mu=0$ it goes back to Qin-Chang model, therefore all hadron properties are preserved.

In the following calculation, we will choose  $D=1.0 \, \textrm{GeV}^2,\  \omega=0.6 \, \textrm{GeV} $ and $m_u=m_d=0.005 \, \textrm{GeV}$ in Eq.~(\ref{eq:QCmodel2}) for demonstration in almost all figures. As for the parameter $\Lambda$,  from Eq.~(\ref{eq:Dk2O2})  we know that it characters the strength of quark's feedback on gluon: the larger $\Lambda$ is, the less quark contributes. When $\Lambda \rightarrow +\infty$, Eq.~(\ref{eq:Dk2O2}) becomes:

\begin{eqnarray}
\label{eq:Dk2O2s}
\hspace{-10mm} g^2 D(\vec{k}^2+\Omega_l^2)\overset{\Lambda \rightarrow + \infty}{\xlongequal{\quad}}g^2 D_s(\vec{k}^2+\Omega_l^2)=\mathcal{G}(\vec{k}^2+\Omega_l^2).
\end{eqnarray}
Here we add a subscript \emph{s} for \emph{static} to this special case for later use. To determine $\Lambda$,  we try to infer its value by comparing with existing study. For example, under Rainbow truncation, \cite{Fischer:2003rp} suggests about $20\%$ increase in $-\langle \bar{\psi}\psi \rangle_{\textrm{u/d}}$ with the unquenching effect. In our case, we have $\langle \bar{\psi}\psi \rangle_{\textrm{u/d}}=-(244 \textrm{MeV})^3$ obtained from $D_{\mu \nu}(k)$ comparing with $\langle \bar{\psi}\psi \rangle_{\textrm{u/d}}^{\textrm{q}}=-(227 \textrm{MeV})^3$ from $D^{\textrm{q}}_{\mu \nu}(k)$ by setting $\Lambda=0.56 \, \textrm{GeV}$.\footnote{$-\langle \bar{\psi}\psi \rangle_{\textrm{u/d}}=(244 \textrm{MeV})^3$ satisfies $-(m_u+m_d) \langle \bar{\psi} \psi \rangle_{\textrm{u/d}} \approx m^2_{\pi}f^2_{\pi}$ (GMOR relation \cite{GellMann:1968rz}) within our parameter setting, although it is relatively small comparing with current lattice prediction $-\langle \bar{\psi}\psi \rangle_{\textrm{u/d}}\approx(270 \textrm{MeV})^3$ \cite{Aoki:2013ldr}.} We will show the responses of CEP location and pseudo-critical line to these parameters at the end of Sec.~\ref{sec:tu}.

\section{Partial restoration of chiral symmetry at $T=0, \mu \ne 0$}
\label{sec:t0}

The research on QCD at zero temperature and finite density is abundant and  causes a lot of interests to researchers on cold QCD matter, e.g., compact stars \cite{Alford:2006vz,Weissenborn:2011qu}. For example, the equation of state (EOS) of cold QCD matter plays an important role in calculating and understanding the structure and evolution of these stars \cite{Buballa:2003qv,Andersen:2002jz,Chen:2012zx,Li:2010gx,Zhao:2015rta}. There have also been studies suggesting a first order phase transition of chiral symmetry on the $\mu$ axis \cite{Asakawa:1989bq,Hell:2011ic}. It is therefore interesting to see the picture from our model.

To solve the quark gap equation, we take the limit $T \rightarrow 0$,  then Eq.~(\ref{eq:gapeq}) becomes
\begin{eqnarray}
\label{eq:gapeqT0}
\hspace{-5mm} [G(\vec{p},\tilde{p}_4)]^{-1}&=&[G^0(\vec{p},\tilde{p}_4)]^{-1}+\int\frac{d^4q}{(2\pi)^4}
\nonumber\\
&&\hspace*{-8mm}\times \, \left[g^2D_{\mu \nu}(\vec{p}\!-\!\vec{q},\tilde{p}_4\!\!-\!\tilde{q}_4)\frac{\lambda^a}{2}\gamma_{\mu}G(\vec{q},\tilde{q}_4)\Gamma_{\nu}^a \right],
\end{eqnarray}
where $\tilde{p}_4$=$p_4+i\mu$, accordingly, the quark propagator can be decomposed as,
\begin{eqnarray}
\label{eq:Gpp4mu}
\hspace{-8mm} G^{-1}(\vec{p},\tilde{p}_4;\mu)&=&i\vec{\gamma}\cdot\vec{p}A(\vec{p}^{~\!2},\tilde{p}_4;\mu)\nonumber \\
&+&i\gamma_4\tilde{p}_4C(\vec{p}^{~\!2},\tilde{p}_4;\mu)
+B(\vec{p}^{~\!2},\tilde{p}_4;\mu).
\end{eqnarray}
We can also calculate the renormalized quark condensate with 
\begin{align}
\label{eq:condT0}
 \langle \bar{\psi} \psi \rangle& = - \int \frac{dp^4}{(2 \pi)^4}\textrm{Tr}_{\textrm{f,c,d}}\left[ G(\vec{p},\tilde{p}_4;\mu)- G^0(\vec{p},\tilde{p}_4;\mu)\right] \nonumber \\
&=-4N_c N_f \int \frac{dp^4}{(2 \pi)^4} \frac{B}{A^2 \vec{p}^2+C^2 \tilde{p_4}^{\!2}+B^2},
\end{align}
where the trace should be taken over flavor, color and Dirac indices assuming the $u$,$d$ quark symmetry.

Substitute Eqs.~(\ref{eq:Dk2O2},\ref{eq:Gpp4mu},\ref{eq:condT0}) into Eq.~({\ref{eq:gapeqT0}}), multiply both sides of  Eq.~(\ref{eq:gapeqT0}) with $i \vec{\gamma}\cdot\vec{p}$, $i \gamma_4 \tilde{p}_4$ and $I_4$ respectively, and then take the trace, one can obtain three coupled nonlinear equations of the functions A, B and C. These nonlinear equations can be numerically solved with iterative method. In this way, we can obtain the scalar functions $A, B$, $C$ and the corresponding quark condensate.

\begin{figure}
\includegraphics[width=0.4\textwidth]{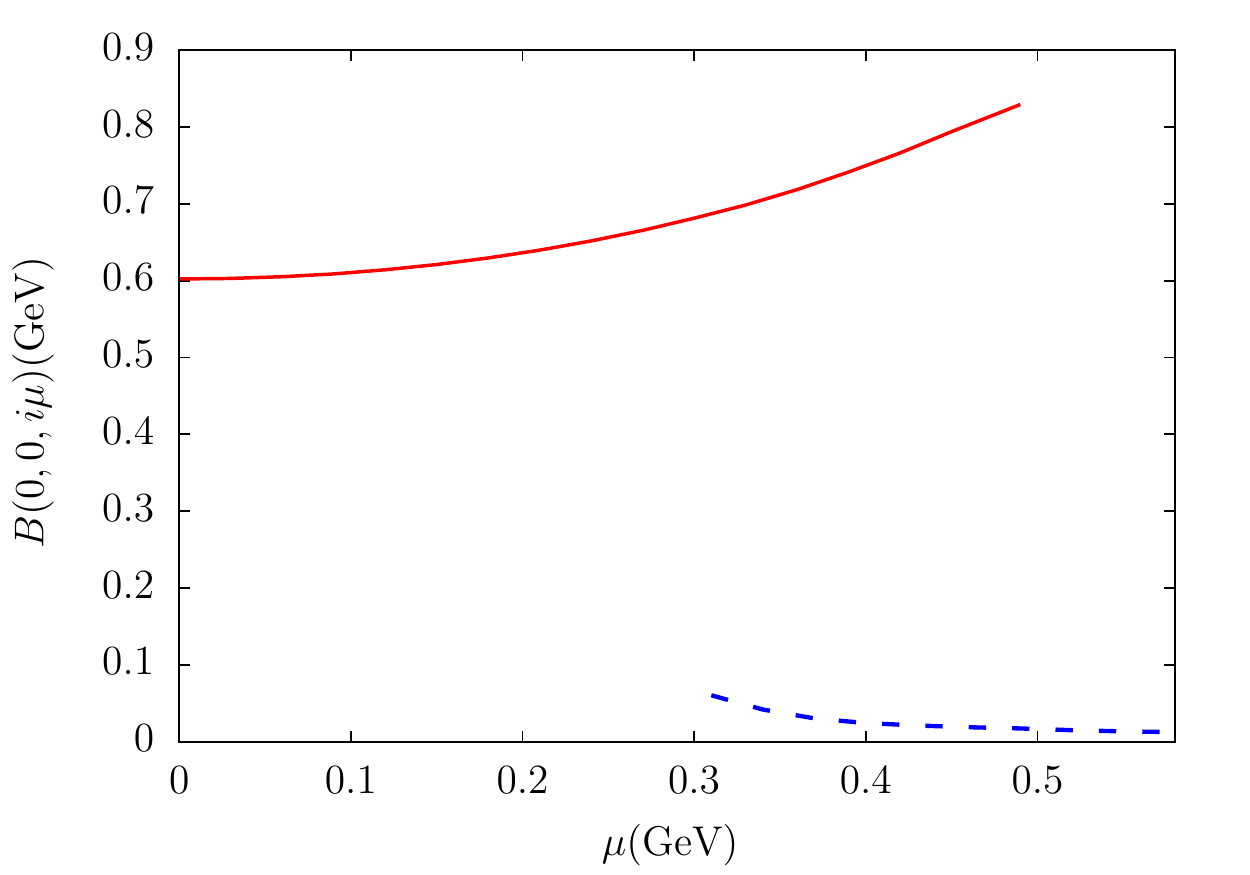}
\caption{Solution of quark gap equation at zero temperature: $B(\vec{p}^{\, 2}, p_4, i \mu)$ with $\vec{p}^{\, 2}=0$, $p_4=0$. Two solutions correspond to the Nambu-Goldstone solution (red solid curve) and Wigner solution (blue dashed curve) respectively.} 
\label{fig:t0B}
\end{figure}

\begin{figure}
\includegraphics[width=0.4\textwidth]{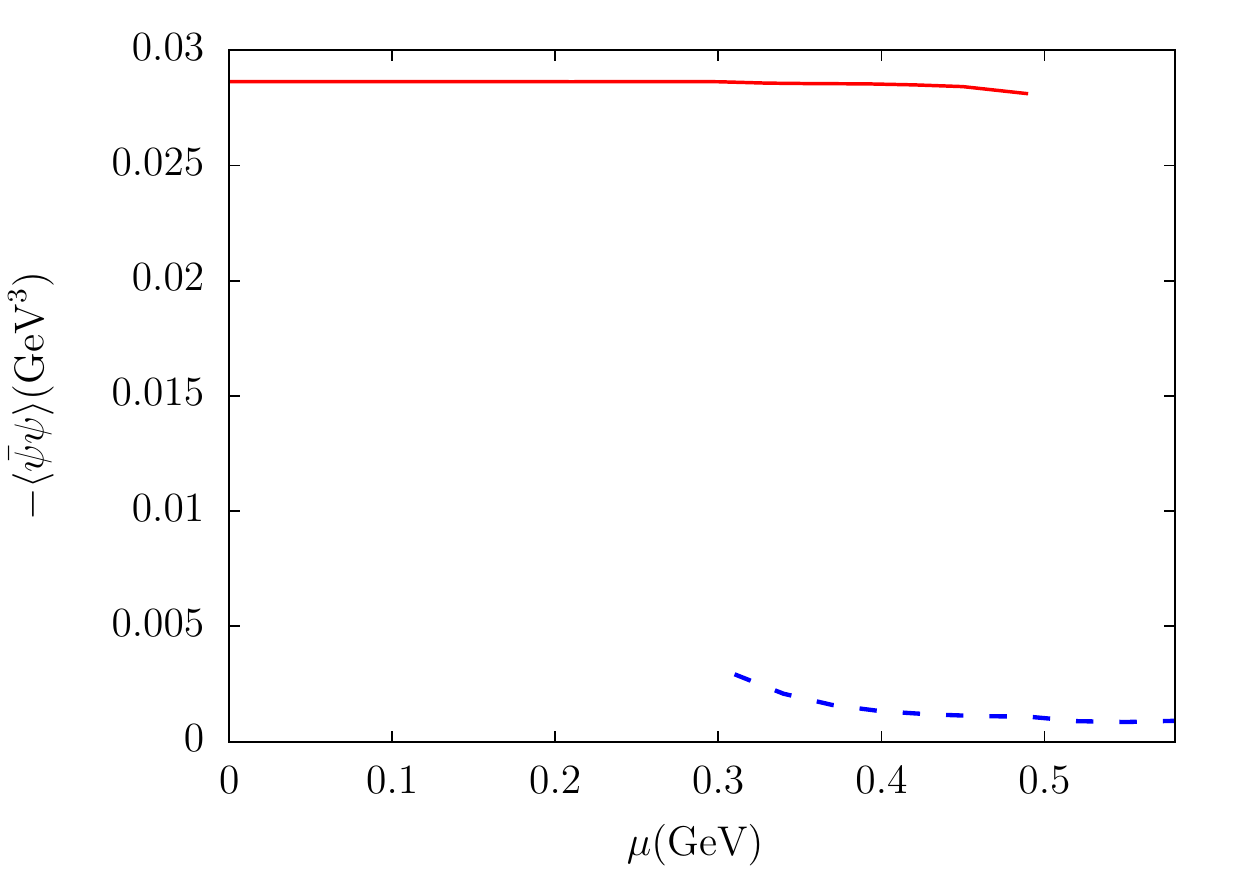}
\caption{Quark condenstate in Nambu-Goldstone phase (red solid curve) and Wigner phase (blue dashed curve).}
\label{fig:t0cond}
\end{figure}

 Both Nambu-Goldstone solution and Wigner solution are found, corresponding to Nambu-Goldstone phase and Wigner phase respectively.  Fig.~\ref{fig:t0B} shows the $B(\vec{p}^{\, 2} \!=0, p_4=0, i \mu)$  for both solutions and Fig.~\ref{fig:t0cond} displays the quark condensates. Both quantities are indicators of DCSB and exhibit discontinuous drop at the same chemical potential. 

One can notice from Fig.~\ref{fig:t0cond} that the $\langle \bar{\psi} \psi \rangle$ in the Nambu phase basically remains unchanged. This indicates that the partition function of QCD stays unchanged before $\mu$ reaches a critical value (roughly 1/3 of baryon mass) \cite{Halasz:1998qr}. Here we'd like to point out this condition can be used as a rule in constraining gluon propagator models, which however was not satisfied at all times. For instance, in Refs.~\cite{Chen:2011my,Jiang:2013xwa,Shi:2014zpa}, a chemical potential suppressed gluon propagator model is employed to study the QCD  phase diagram. While  $\langle \bar{\psi} \psi \rangle$ in that case actually varies with chemical potential and therefore breaks this rule to a certain extent. In our model here, this condition is satisfied because the whole quark's feedback is incorporated into a term solely described by quark condensate, which was already unchanged in the Nambu-Goldstone phase with static gluon models.

To study the possible phase transition between two phases, one should calculate the effective potential and obtain the pressure difference between them, whose zero point at $\mu_c$  is where the first order phase transition takes place. However, the Cornwall-Jackiw-Tomboulis (CJT) effective potential action could only be used consistently with the rainbow truncation and a static gluon propagator model \cite{Cornwall:1974vz}, thus is invalid here. Nevertheless, suggested by other studies beyond Rainbow truncation, e.g.  the Ball-Chiu vertex,  first order phase transition should take place within the coexistence region of two solutions \cite{Chen:2008zr, Qin:2010nq}. An intuitive guess for the first order phase transition point is \cite{Chen:2008zr}
\begin{eqnarray}
\mu_c^{\chi} \approx \frac{\mu_c^{NG}+\mu_c^{W}}{2}=0.4 \textrm{GeV},
\end{eqnarray}
 $\mu_c^{NG}$ is where Nambu-Goldstone solution disappears while $\mu_c^{W}$ is where Wigner solution turns up. We'd also like to point out the $\mu_c^{W}$ from $D(k^2)$  is about 20 MeV lower than that from $D_s(k^2)$, indicating a small decrease in $\mu_c^{\chi}$ within our model.

Finally, one could infer from Fig.~\ref{fig:t0cond} and Eq.~(\ref{eq:Dk2O2}) that our gluon propagator takes different forms in Nambu-Goldstone phase and Wigner phase,  $D_{NG}(k^2) = D_s(k^2)$ comparing with $D_{W}(k^2) \simeq  D^q(k^2)$. Therefore, our gluon propagator has a clear distinction between the Nambu-Goldstone phase and Wigner phase, which gives a solution to the problem we proposed in Sec.~\ref{sec:intro}. In this way, not only quark, but also gluon propagators take discontinuous changes while the system goes through a first order phase transition. This gives a general picture about how the gluon propagator evolves at finite $\mu$, through the inclusion of quark's feedback.

\section{partial restoration of chiral symmetry at $T \ne 0$}
\label{sec:tu}

We now move on to the finite temperature case and solve the gap equation at finite $T$ and $\mu$. The quark condensate at finite temperature is
\begin{eqnarray}
\label{eq:condT}
 \langle \bar{\psi} \psi \rangle = -T \sum_{n=-\infty}^{+\infty} \int \frac{dp^3}{(2 \pi)^3}\textrm{Tr}_{\textrm{f,c,d}}[ G(\vec{p},\tilde{\omega}_n;T,\mu)\nonumber \\
- G^0(\vec{p},\tilde{\omega}_n;T,\mu)].
\end{eqnarray}
 Taking the limit $T\rightarrow 0$ in this equation leads  
 to Eq.~(\ref{eq:condT0}). Following similar steps as introduced in Sec.~\ref{sec:t0} (replace $\tilde{p}_4$ with $\tilde{\omega}_4$), we can again obtain the dressing functions and corresponding quark condensate. 

\begin{figure}
\includegraphics[width=0.43\textwidth]{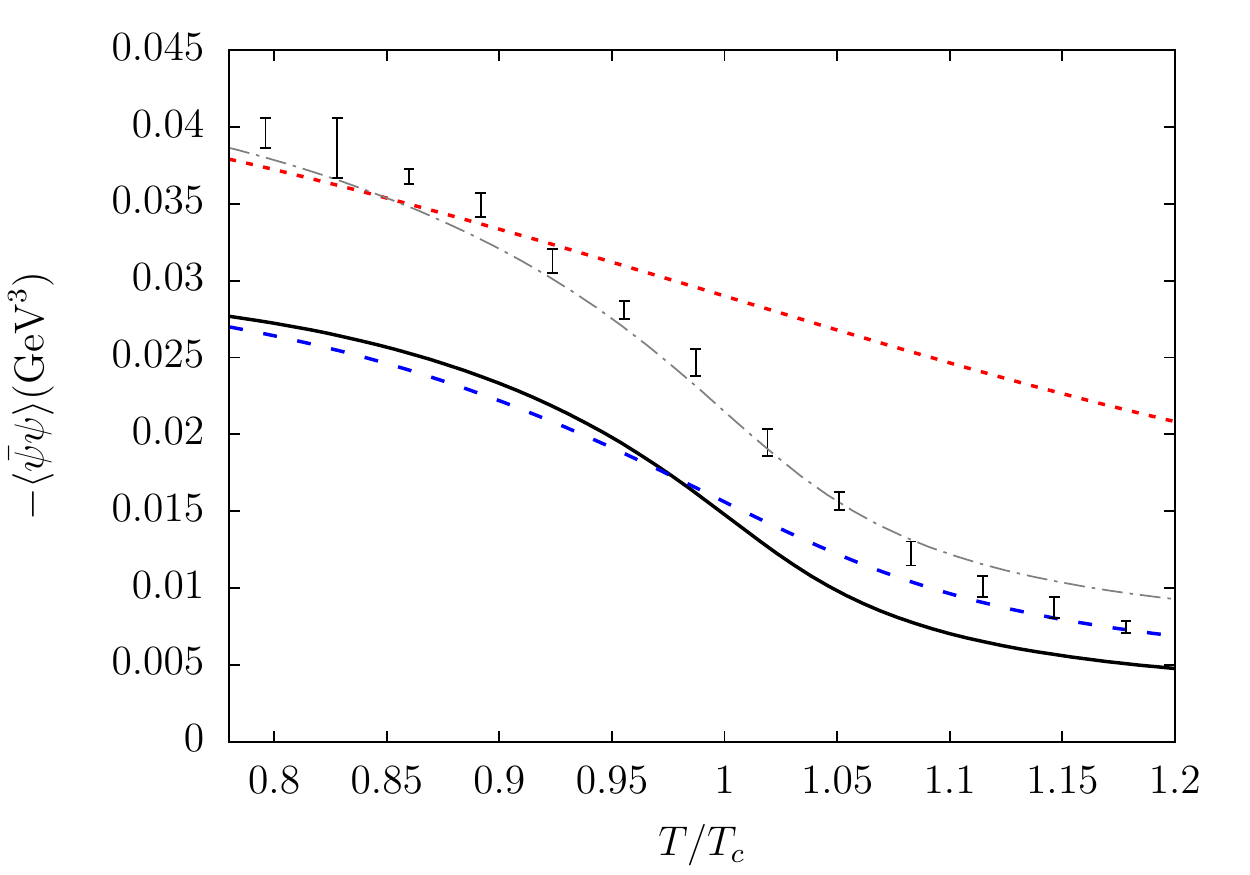}
\caption{Evolution of $\langle \bar{\psi} \psi \rangle$ at $T\ne0$ and $\mu=0$. $\langle \bar{\psi} \psi \rangle$ (black solid curve) from our modified model Eq.~(\ref{eq:Dk2O2}) decreases faster than $\langle \bar{\psi} \psi \rangle_s$ (blue dashed curve) from the static model Eq.~(\ref{eq:Dk2O2s}). The other $\langle \bar{\psi} \psi \rangle$ (gray dot-dashed curve) and $\langle \bar{\psi} \psi \rangle_s$ (red dotted curve) were obtained by setting the parameter $D=1.4$ and $\Lambda=0.62$, whose choice is explained in the text. The data (black error bars) are taken from lattice calculation \cite{Borsanyi:2010bp}.}
\label{fig:u0cond}
\end{figure}

\begin{figure}
\includegraphics[width=0.4\textwidth]{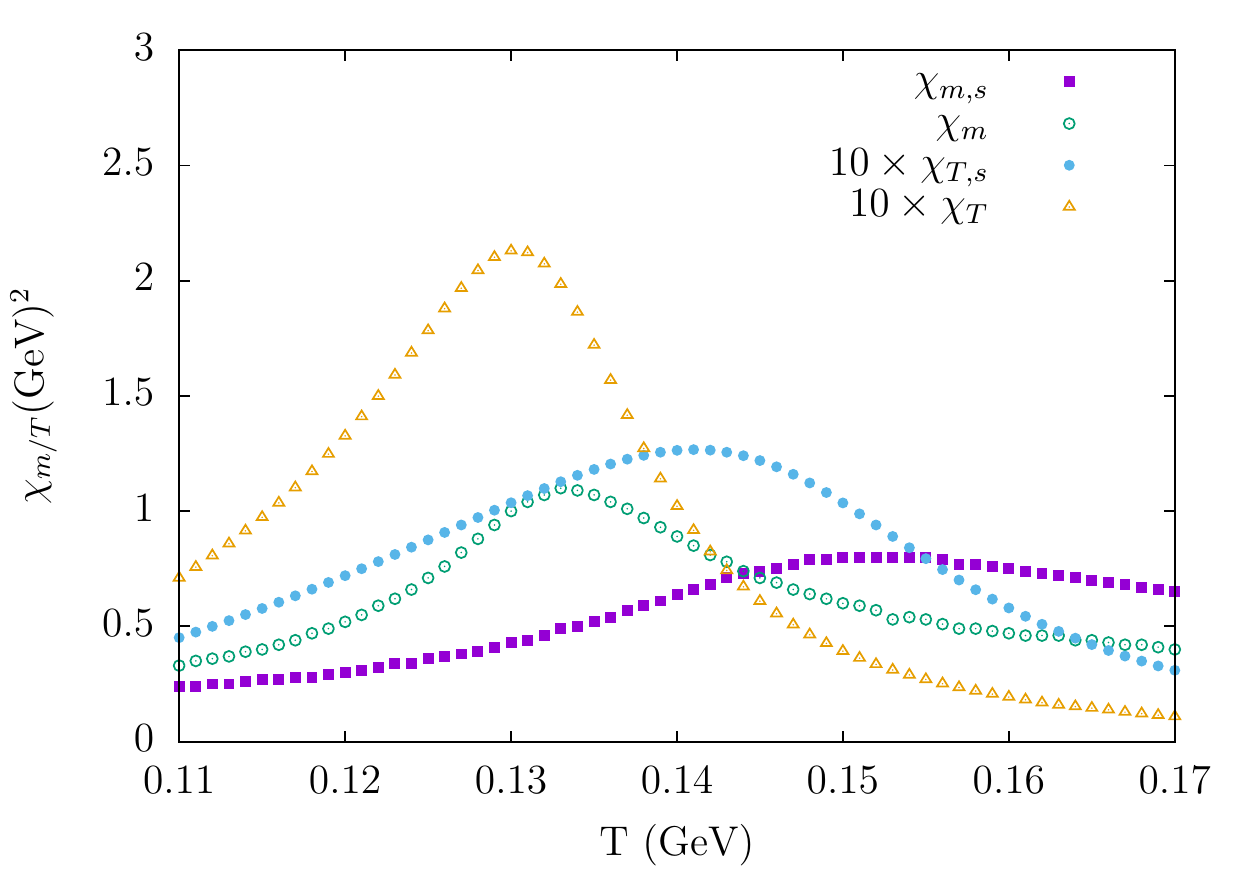}
\caption{Susceptibilities at finite temperature and zero chemical potential. $\chi_{T}$ and $\chi_m$ are defined in Eq.~(\ref{eq:chiT}) and Eq.~(\ref{eq:chim}) with the subscript \emph{s} for \emph{static}.}
\label{fig:u0sus}
\end{figure}

Let's first look at the results on the temperature axis, namely $\mu=0$. As we can see from Fig.~\ref{fig:u0cond}, introducing quark's feedback doesn't change the qualitative behavior of quark condensate on the $T$ axis. $\langle \bar{\psi} \psi \rangle$ is basically a monotonic decreasing function of $T$ with an inflection point. If we use the susceptibility 
\begin{eqnarray}
\label{eq:chiT}
\chi_T=\frac{\partial \langle \bar{\psi} \psi \rangle}{\partial T}
\end{eqnarray}
 as the criterion \cite{Morita:2011jva,Skokov:2011ib}, this inflection point is the so called pseudo-critical temperature. Another choice is the chiral susceptibility $\chi_m$
\begin{eqnarray}
\label{eq:chim}
\chi_m=-\frac{\partial \langle \bar{\psi} \psi \rangle}{\partial m}.
\end{eqnarray}
The maxima of $\chi_T$ and $\chi_m$, namely the pseudo-critical temperatures---$T_c$'s, don't necessarily coincide with each other \cite{Du:2013oza}, although within our model they are closer (see Fig.~\ref{fig:u0sus}). Nevertheless, all these susceptibilities exhibit smooth change hence it is crossover in this area. 

However, some quantitative changes are noticeable. In Fig.~\ref{fig:u0sus}, $T_c$'s from two gluon propagator models are not the same: $D(k^2)$ gives a relatively low $T_c$. This can be understood with the help of Fig.~\ref{fig:u0cond}: when $T$ goes up,  $-\langle \bar{\psi} \psi \rangle$ drops continuously, so $D(k^2)$ gets smaller and leads to the weakening of interaction between quarks. This then in turn accelerates the dropping of quark condensate, producing relatively low $T_c$'s. Comparing with that of static gluon model, this ``quicker'' transition brought by quark's feedback is closer to lattice result concerning the slopes of $\langle \bar{\psi} \psi \rangle$ curves near $T_c$. We notice the $\langle \bar{\psi} \psi \rangle$ and $T_c$'s we give are relatively low comparing to lattice predictions  $T_c\approx 160$ MeV and $\langle \bar{\psi} \psi \rangle_{\textrm{u/d}} \approx 280$  MeV given in \cite{Borsanyi:2010bp}. This is due to our simplified truncation scheme and gluon model. In order to perform a direct comparison with lattice result, we primitively raise the interaction strength to $D=1.4$ and $\Lambda=0.62$ (the choice of $\Lambda$ is explained in the caption of Table.~\ref{tab:Lambda}), which produce $\langle \bar{\psi} \psi \rangle_{\textrm{u/d}} = 277$ MeV. The $\langle \bar{\psi} \psi \rangle_T$ still agrees with lattice data better than a static model, as shown in Fig.~\ref{fig:u0cond}. Therefore we conclude that our modified model persists to give more realistic descriptions of transition behavior at finite temperature $T \leqslant  T_c$.

 It is worth noting that, in contrast with that on $\mu$ axis where $D(k^2)$ takes a sudden change,  here $D(k^2)$ decreases continuously. Such a behavior is naturally generated in our model and qualitatively agrees with lattice result on $D_T(k^2)$ \cite{Fischer:2010fx}.  
 
\begin{figure}
\includegraphics[width=0.5\textwidth]{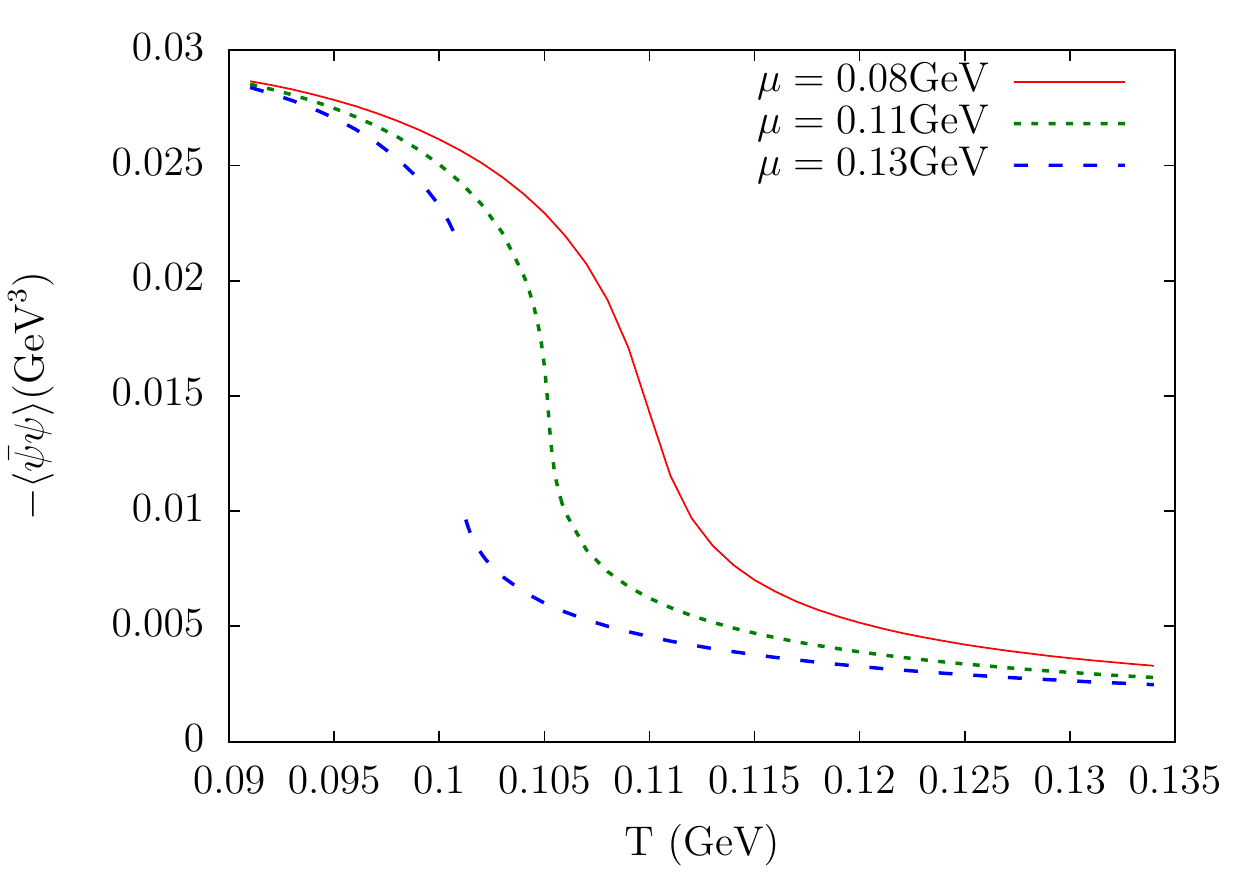}
\caption{$\langle \bar{\psi} \psi \rangle$ at finite $\mu$ and $T$.}
\label{fig:tucond}
\end{figure}

\begin{figure}
\includegraphics[width=0.5\textwidth]{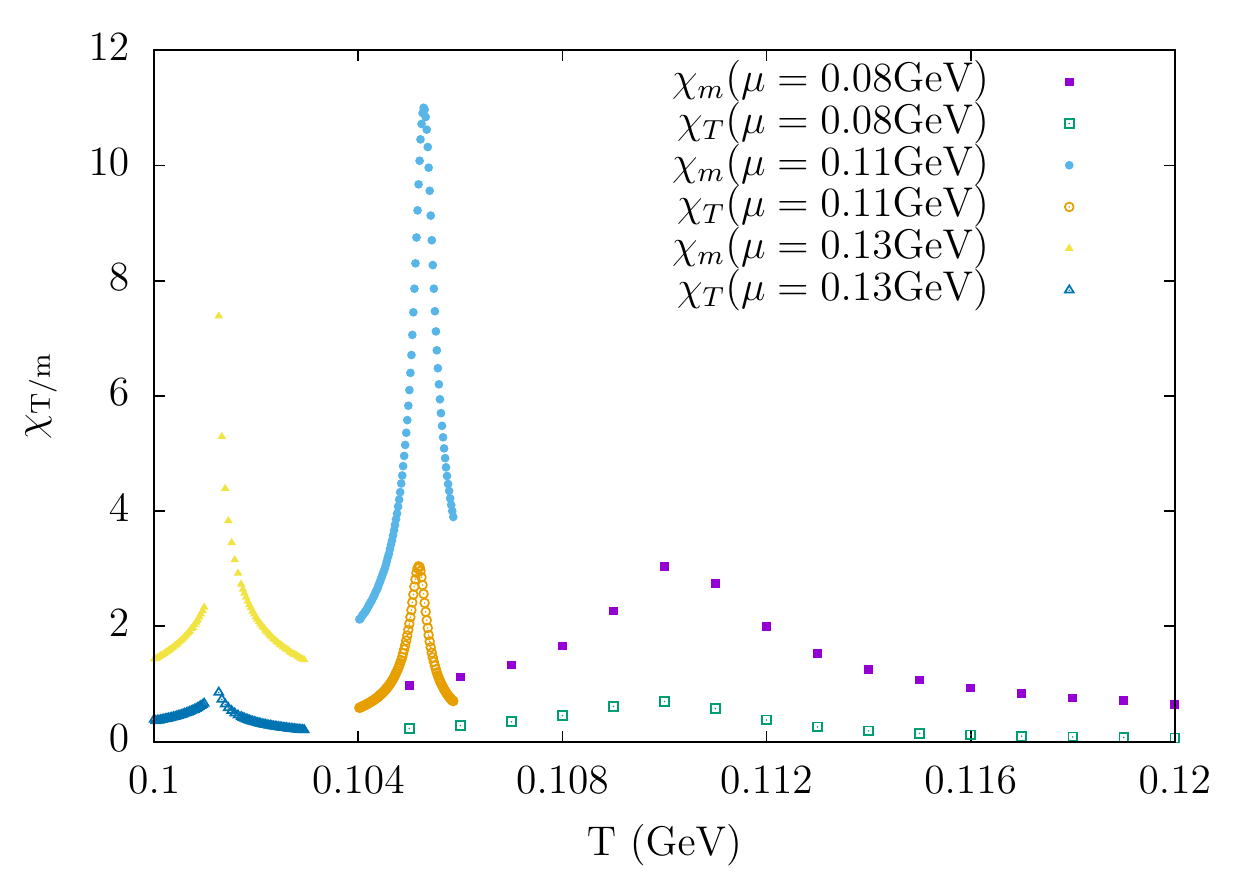}
\caption{Crossover and first order phase transition characterized by susceptibilities. The peaks go to infinity approaching the critical end point.}
\label{fig:tusus}
\end{figure}

 With the results on $\mu-$axis and $T-$axis, we are led to believe transition behaviors like crossover and first  order phase transition will remain on $T-\mu$ plane, while the transition lines will somehow vary. Consequently, the CEP, which is the end point of first order phase transition line, may shift. 


$\langle \bar{\psi} \psi \rangle$ at $T \ne 0$ and $\mu \ne 0$ are shown in Fig.~\ref{fig:tucond}, where $\langle \bar{\psi} \psi \rangle$ undergoes continuous change with low $\mu$ while exhibits discontinuous  transition with larger $\mu$. Fig.~\ref{fig:tusus} shows that the corresponding susceptibilities display different behaviors, e.g.,  at $\mu=110$ MeV and $\mu=80$ MeV they are continuous while at $\mu=130$ MeV not. One could also see there is a tendency for the susceptibilities to diverge at some point when $\mu$ is larger than $110$ MeV. This point is then the second order phase transition point, namely, the CEP. 

\begin{figure}[htbp]
\includegraphics[width=0.5\textwidth]{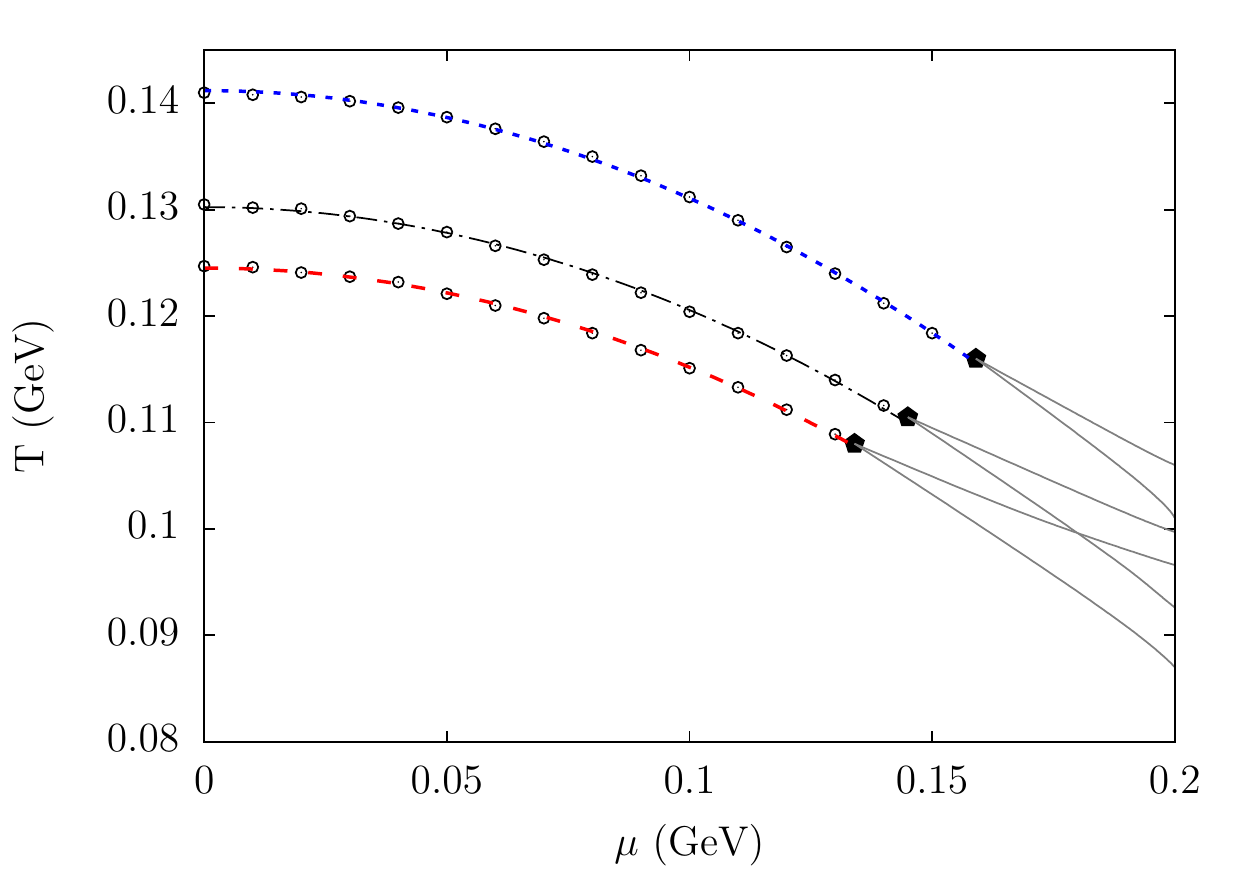}
\caption{From top to bottom, pseudo-transition points obtained from $\Lambda\!\!=\!\!+\infty \textrm{ (blue dotted curve)},  \Lambda=0.56$ (black dot-dashed line) and $\Lambda=0.5$ (red dashed curve) respectively, are all well fitted by  Eq.~(\ref{eq:kappa}). The area between the gray solid curves is the metastable region of Nambu-Goldstone phase and Wigner phase.}
\label{fig:kappa}
\end{figure}

We therefore determine the pseudo-critical lines in Fig.~\ref{fig:kappa} by taking the maxima of $\chi_T$. For comparison, again we also add the result of $D_s(k^2)$. It shows, with quark's feedback on gluon propagator, the pseudo-critical line gets flattened.   This can be seen more clearly with the parameterization formula \cite{Kaczmarek:2011zz,Endrodi:2011gv}:

\begin{eqnarray}
\label{eq:kappa}
T_c(\mu)=T_c(0)(1-\kappa \mu^2/T_c^2(0)+O(\mu^4/T_c^4)),
\end{eqnarray}
where $T_c(\mu)$ parameterizes the pseudo-transition line. We extract $\kappa$ by least-squares fit  and plot the functions $T_c(\mu)$'s  in Fig.~\ref{fig:kappa}. The root-mean-square deviation in this fitting,
\begin{align}
\label{eq:sd}
\textrm{RMSD}=\sqrt{\frac{1}{N} \sum_{i=1}^{N}\biggl (T_c(\mu_i)-T_c^i\biggl )^2}
\end{align}
is $\textrm{RMSD} < 0.2$ MeV for all curves.

\begin{table}[htbp]
\centering
\begin{tabular}{cccccc}
\hline
\hline
D & $\omega$&$\Lambda$ (GeV) & \ \ $T_c$ (GeV) & \ \ $(T_E,\mu_E)/T_c$ & \hspace{5mm} $\kappa$ \hspace{5mm} \\
\hline
 1.0 &\  0.6&0.5&0.125&(0.89,1.01)&0.116 \\
 1.0 &\  0.6&0.56&0.131&(0.85,1.11)& 0.126\\
 1.0 &\  0.6& $+\infty$&0.141  &(0.82,1.13)&0.143 \\
 1.0 &\  0.5&0.52&0.156&(0.93,0.41)&0.333 \\
 1.4 &\  0.6&0.62&0.176&(0.93,0.46)&0.323 \\
\hline
\hline
\end{tabular}
\caption{\label{tab:Lambda} Parameter dependence of the CEP location and curvature parameter $\kappa$ defined in Eq.~(\ref{eq:kappa}). In row-4 and row-5, $\Lambda$ is determined by the same criterion as in row-2: $\langle \bar{\psi} \psi \rangle_0^\textrm{q} / \langle \bar{\psi} \psi \rangle_0=0.8$ .}
\end{table}

The first three rows in Table. \ref{tab:Lambda} show $\kappa$'s from different $\Lambda$'s, along with the $T_c$'s and CEP locations in Fig.~\ref{fig:kappa}. we can see there is consistent decrease in $\kappa$ and increase in $T_E/T_c$ as $\Lambda$ decreases. Same conclusion can be drawn when we employ the Maris-Tandy model, namely Eq.~(\ref{eq:MTmodel}), for which the calculation will not be detailed here. Note that lattice QCD suggests $\kappa \approx 0.05-0.06$ \cite{Kaczmarek:2011zz,deForcrand:2010he,Endrodi:2011gv} and estimates $(T_E,\mu_E)/T_c \approx(0.9-0.95,1.0-1.4)$ \cite{Sharma:2013hsa} . Given that in general, model studies tend to give relatively large $\kappa$ and low $T_E/T_c$ \cite{Stephanov:2007fk}, our model therefore provides a means for improvements in these cases. 

In the last two rows of Table. \ref{tab:Lambda}, the response of the CEP's location to varying the parameters $D$ and $\omega$ is shown respectively. As explained in the end of \cite{Qin:2010nq}, if we consider $r=1/ \omega$ as a confinement length scale, then when $r$ goes to zero, which represents a NJL type model, the CEP would rotate toward the chemical potential axis. Therefore the CEP rotates toward the temperature axis from row-2 to row-4.  In Row-5, the interaction strength $D$ is raised to produce a larger $-\langle \bar{\psi} \psi \rangle_{\textrm{u/d}} \approx (280 \textrm{MeV})^3$, which corresponds to the gray dot-dashed curve in Sec.~~\ref{fig:u0cond}. CEP in this case also rotates toward the temperature axis. So generally speaking,  reducing $\omega$ would make CEP rotate toward the temperature axis under constraint of $D \omega \approx (0.8 \textrm{GeV})^3$.

\section{Discussion and Summary}\label{sec:summary}

To summarize, we incorporate quark's feedback into the gluon propagator based on the idea of OPE and derive a gluon propagator that evolves through the $T-\mu$ plane. It is characterized and determined by quark condensate at finite temperature and density.  The QCD  phase diagram is then studied with this gluon model within DSEs framework.

 At zero temperature and finite chemical potential,  the coexistence region of Nambu-Goldstone solution and Wigner solution is found, indicating a first order phase transition point. Moreover, we have shown that our model preserves two important features of QCD, e.g., QCD remains vacuum at low chemical potential and discontinuous change in gluon propagator at the first order phase transition. Then we move on to $T \ne 0$ case and find that quark's feedback accelerates the decrease of quark condensate, leading to a quicker crossover on the temperature axis. Such a picture agrees with lattice simulation at finite temperature. We further studied the crossover region and  CEP location. It shows consistent decrease in curvature parameter $\kappa$ and increase in $T_E/T_c$ with more of gluon propagator distributed to quark's feedback. For example, it brings a CEP location from $(T_E,\mu_E)/T_c =(0.82,1.13)$ to $(T_E,\mu_E)/T_c =(0.85,1.11)$ and $\kappa$ from $0.143$ to $0.126$, both closer to lattice estimation. We therefore believe our scheme could provide a means for improvements in model studies which haven't considered quark's feedback.

  Finally, it is worth noting that this work is a supplement to existing investigation with refined quark-gluon vertexes beyond Rainbow truncation. Authors of \cite{Qin:2010nq} have shown that with the Ball-Chiu vertex, QCD  phase diagram is improved in several aspects, e.g., significantly narrower metastable region and more reasonable CEP location. Since the dressing effect in Ball-Chiu vertex is also expressed in terms of quark's dressing functions and therefore consists of quark's feedback, it's evident that the incorporation of quark's feedback within DSEs framework could produce QCD phase diagram that is more realistic.

\acknowledgments

We benefit from discussion with Fei Gao and thank him for providing valuable insights. This work is supported in part by the National Natural Science Foundation of China (under Grant Nos. 11275097, 11475085, 11265017, and 11247219), the National Basic Research Program of China (under Grant No. 2012CB921504), the Jiangsu Planned Projects for Postdoctoral Research Funds (under Grant No. 1402006C), the National Natural Science Foundation of Jiangsu Province of China (under Grant No. BK20130078), and Guizhou province outstanding youth science and technology talent cultivation object special funds (under Grant No. QKHRZ(2013)28).

\bibliography{DG11756}

\end{document}